\documentclass[numberedappendix, twocolappendix]{emulateapj}

\usepackage{natbib} 
\usepackage{times}
\usepackage{color}
\usepackage{graphicx}
\usepackage{cases}
\definecolor{RoyalBlue}{rgb}{0.25,0.41,0.88}

\shorttitle{Dust Trap in the MWC58 disk} \shortauthors{Marino et al.} 

\begin{document}

\title{Compact dust concentration in the MWC~758 protoplanetary disk}

\author{S. Marino\altaffilmark{1,2}, S. Casassus\altaffilmark{1,2},
  S. Perez\altaffilmark{1,2}, W. Lyra\altaffilmark{3},
  P.E. Roman\altaffilmark{4,2}, H. Avenhaus\altaffilmark{1,2},
  C.M. Wright\altaffilmark{5} and S.T. Maddison\altaffilmark{6}.}

\affil{$^1$Departamento de Astronom\'ia, Universidad de Chile,
  Casilla 36-D, Santiago, Chile}

\affil{$^2$Millenium Nucleus ``Protoplanetary Disks in ALMA Early
  Science,'' Universidad de Chile, Casilla 36-D, Santiago, Chile}

\affil{$^3$Department of Physics and Astronomy, California State
  University Northridge, 18111 Nordhoff St, Northridge, CA 91330, USA}

\affil{$^4$Center of Mathematical Modeling, Universidad de Chile,
  Beauchef 851, Santiago, Chile}

\affil{$^5$School of Physical, Environmental and Mathematical
  Sciences, UNSW@ADFA, Canberra ACT 2600, Australia}

\affil{$^6$Centre for Astrophysics and Supercomputing, Swinburne
  University of Technology, PO Box 218, Hawthorn, VIC 3122, Australia}

\email{smarino@das.uchile.cl}
  

\begin{abstract}
  
The formation of planetesimals requires that primordial dust grains
grow from micron- to km-sized bodies. Dust traps caused by gas
pressure maxima have been proposed as regions where grains can
concentrate and grow fast enough to form planetesimals, before
radially migrating onto the star. We report new VLA Ka \& Ku
observations of the protoplanetary disk around the Herbig Ae/Be star
MWC 758. The Ka image shows a compact emission region in the outer
disk indicating a strong concentration of big dust grains. Tracing
smaller grains, archival ALMA data in band 7 continuum shows extended
disk emission with an intensity maximum to the north-west of the
central star, which matches the VLA clump position. The compactness of
the Ka emission is expected in the context of dust trapping, as big
grains are trapped more easily than smaller grains in gas pressure
maxima. We develop a non-axisymmetric parametric model inspired by a
steady state vortex solution with parameters adequately
  selected to reproduce the observations, including the spectral
energy distribution. Finally, we compare the radio continuum with
SPHERE scattered light data. The ALMA continuum spatially coincides
with a spiral-like feature seen in scattered light, while the VLA
clump is offset from the scattered light maximum. Moreover, the ALMA
map shows a decrement that matches a region devoid of scattered
polarised emission. Continuum observations at a different wavelength
are necessary to conclude if the VLA-ALMA difference is an opacity or
a real dust segregation.
\end{abstract}

\keywords{protoplanetary disks - planet–disk interactions - techniques: interferometric}

\section{Introduction}

The formation of terrestial planets or of giant planets by core
accretion requires the build up of a population of planetesimals. To
form these bodies, dust grains need to grow from micron to kilometer
sizes, i.e. grains need to increase in size by at least nine orders of
magnitude in size scale, until the influence of gravity enhances the
protoplanet growth \citep{Safronov1969, Pollack1996}. This process is
a race against the disk dispersal process \citep{Hollenbach1994,
  Clarke2001} and accretion onto the central star
\citep{Hartmann1998}. 


As dust grains grow, they start to decouple from the gas and drift
radially towards the central star
\citep{Weidenschilling1977MNRAS.180...57W}. Therefore, planetesimal
formation has to happen in time scales shorter than radial drift. One
way to halt the inward drift is by developing a local maximum in the
radial surface density of the gas. Dust grains can easily get trapped
in density maxima \citep{Whipple1972, Barge1995, Haghighipour2003, Chiang2010},
especially big grains \citep{Brauer2008, Pinilla2012}. Dust traps can
be found in the inner edges of disks with evidence of gaps and
cavities, commonly called transitional disks
\citep{Espaillat2014}. These gaps and cavities are also detected as
infrared dips in the spectral energy distribution (SED). In these dust
traps, and depending on the level of gas turbulence, grains can
accumulate for periods comparable to the disk lifetime. Thereby
benefiting grain growth, and producing an enhancement in the
population of larger dust grains \citep{Brauer2008, Pinilla2012}.

Azimuthal asymmetries, such as vortices, can also produce dust
trapping \citep{Barge1995, Tanga1996, Johansen2004, Inaba2006,
  Lyra2008, Lyra2009a, Regaly2012, Lyra2013}. Large scale asymmetries
are observed in sub-millimeter observations of the transition disks
around SAO~206426, SR~21 \citep{LPerez2014}, LkH$\alpha$~ 330
\citep{Isella2013}, and particularly in HD~142527
\citep{Casassus2013Natur, Fukagawa2013} and IRS~48
\citep{vanderMarel2013}, where a stronger than usual azimuthal
contrast is observed.  These observations suggest dust trapping by
large-scale vortices. In this scenario it is expected that the dust
spatial distribution will depend strongly on the grain size, as
aerodynamic drag depends on the grain cross section. Big grains will
concentrate in more compact regions than small grains
\citep{Barge1995, Klahr1997, Dubrulle2005, Inaba2006, Meheut2012b,
  Lyra2013, Raettig2015}. This has strong observable implications: at
long wavelengths, dust continuum emission should be more compact than
at shorter wavelengths, as it traces the mass distribution of grains
with sizes comparable to the observed wavelength \citep{Beckwith1990,
  Testi2003, Birnstiel2013}. On the other hand, \cite{Mittal2015} show
that large and similar structures to the observed asymmetries in
protoplanetary disks can be formed by gravitational global
  modes, which are currently difficult to discriminate from
  vortices. Thus, in the absence of tracers of the gas mass, it is
crucial to use multiwavelength observations to diagnose if dust
trapping around a pressure maximum is present.


In this paper we study the transition disk around the Herbig Ae star
MWC~758 (HD~36112). The main parameters for this system are given in
Table \ref{tab:param}. The central star is $3.7\pm2.0$ Myr old
\citep{Meeus2012}, and optical spectroscopy reveals variable profiles
suggesting jet-like inhomogeneities at different distances from the
star \citep{Beskrovnaya1999}. The revised \textit{Hipparcos} parallax
data put the star at a distance of $279^{+94}_{-58}$ pc
\citep{vanLeeuwen2007}, although most previous studies use the old
\textit{Hipparcos} estimated distance of $200^{+60}_{-40}$ pc
\citep{vandenAncker1998}. Studying the $^{12}$CO(3-2) kinematics,
\citet{Isella2010} determined a stellar mass of $2.0 \pm 0.2
M_{\odot}$ and a disk orientation with an inclination ($i$) of
$21\pm2^{\circ}$ and a position angle (PA) of $65 \pm 7^{\circ}$, all
consistent with Keplerian rotation. Despite its bright infrared
excess, a cavity of $0.37''$ (73 astronomical units (au) at 200 pc)
has been inferred from dust millimeter emission suggesting a depletion
of millimeter grains in the inner regions of the disk possibly due to
the presence of a low-mass companion within a radius of 42 au
  (using the revised distance) \citep{Isella2010, Andrews2011}. In
addition, the SED shows a 10 $\mu$m dip consistent with a
pre-transitional disk \citep{Grady2013}, but with a strong
near-infrared emission coming from sub-au scales observed with VLTI
\citep{Isella2008}.

\begin{table}[h]
 \centering
   \caption{Main parameters of MWC~758.} 
   \begin{tabular}{ccc}
    \hline
    \hline
    Parameter & Value  & Reference \footnote{ (1) \citealt{vanLeeuwen2007}; (2) \citealt{Meeus2012}; (3) \citealt{Isella2010} ; (4) \citealt{Hog2000}; (5) \citealt{Cutri2003}; } \\ 
    \hline
    RA  (J2000)       &  $05^{h} \ 30^{m} \ 27.530^{s}$  & (1) \\
    DEC (J2000)       &  $+25^{\circ} \ 19'\  57.082''$  & (1)  \\
    Age               &   $3.7\pm2.0$ Myr                  &   (2) \\
    Stellar Mass      &   $2.0 \pm 0.2 M_{\odot} $       &   (3)  \\
    $V$                 &   8.27                          &   (4)   \\      
    $B-V$               &   0.3                           &   (4)    \\
    $H$                 &   6.56                          &   (5)   \\
    Distance          &   $279^{+94}_{-58}$ pc            &    (1)   \\
    Disk inclination  &   $21 \pm 2^{\circ}$             &   (3)   \\
    Disk PA           &   $65 \pm 7^{\circ}$             &   (3)   \\
    \hline
   \end{tabular} 
\label{tab:param}
\end{table}


The disk structure shows deviations from an axisymmetric disk. Spiral
arms are detected from near-IR scattered light and thermal emission
images extending from the sub-millimeter cavity to the outer disk
\citep{Grady2013, Benisty2015}, possibly produced by a massive
  planet at $\sim$160 au \cite{Dong2015}. The drop in the polarised
intensity beyond the spiral arms can be interpreted as a shadowing
effect by the arms, as the disk extends farther out in the
millimeter. In addition, at millimeter wavelengths a peak intensity in
the dust continuum has been detected to the north-west of the central
star after subtraction of a best-fit axisymmetric model, suggesting an
asymmetry in the millimeter dust grain distribution in the outer parts
of the disk \citep{Isella2010}. Moreover, in the same study,
asymmetries in CO emission were observed that could be due to a warped
optically thick inner disk \citep{Isella2010}, similar to the case of
HD~142527 \citep{Marino2015a}. If the southern spiral arm is trailing,
the disk is rotating with a clockwise direction in the plane of the
sky, where the north side of the disk is the nearest.

We analyse new VLA Ka \& Ku observations and we compare with ALMA
archival data to study the distribution of dust grains and search for
evidence of grain size segregation due to dust trapping. In Section
\ref{sec:data} we describe the data and imaging. In Section
\ref{sec:Analysis} we investigate the origin of the VLA Ka \& Ku
asymmetric emission, comparing with archival ALMA and SPHERE
data. In Section \ref{sec:model} we describe a parametric
asymmetric disk model inspired by a modified vortex solution,
  that can reproduce the basic morphology seen in the VLA and ALMA
images and the SED, to try to interpret the observed asymmetries. In
Section \ref{sec:discussion} we discuss about the origin of the
asymmetries: A large scale pressure maximum, e.g. a vortex
  formed at the outer edge of the cavity producing dust trapping, or
  an embedded massive forming planet accreting material while heating
  its its surroundings. Finally in Section \ref{sec:Conclusions} we
summarise the main conclusions of this work.

\section{Observations and Imaging} \label{sec:data}

In this section we describe the observations and imaging of the VLA Ka
($\sim$33 GHz) \& Ku ($\sim$15 GHz) and ALMA Band 7 ($\sim$337 GHz)
data. The image synthesis was carried out using a non-parametric
least-squares modeling technique with a regularization term called
entropy from the family of maximum entropy methods (MEM). We call the
whole algorithm and code \textit{uvmem} and the resulting deconvolved
model images are labelled as 'MEM model'. Examples of MEM in
  astronomy can be found in \cite{Pantin1996, Casassus2006,
    Levanda2010, Casassus2013Natur, Warmuth2013, Coughlan2013}.  It
is possible to characterize the resolution level with a synthetic beam
derived from the response of the algorithm to a point source with the
same noise level as the observations. The MEM synthetic beam is
usually finer than the standard Clean beam with natural or
\textit{briggs} weighting.


The deconvolved model images are 'restored' by adding the dirty map of
the residuals of uvmem and convolving with a Clean beam
characteristic of natural or briggs weighting. These
images can be compared with Clean images and are labelled as 'restored
images'.

\subsection{VLA}

The VLA observations form part of the project VLA 13B-273. The MWC~758
observations were acquired on 6 runs: 3 nights in October and November
of 2013 in Ka band and another 3 nights in January of 2014 in Ku
band. In Ka band the target was observed in B 0 configuration with
minimum and maximum projected baselines of 190 m and 10.5 km, while in
Ku band it was observed in BnA 0 configuration with minimum and
maximum projected baselines of 80 m and 23 km. The array included 27
antennas of 25 m diameter, and the total integration time on source
was 1h 21min in each band.

For Ka band, the VLA correlator was configured to produce 64 spectral
windows from 28.976 GHz to 37.024 GHz, with a spectral window
bandwidth of 128 MHz, each divided into 64 channels to study the
continuum emission with a resolution of 2 MHz. In Ku band, the
correlator was configured similarly with 64 spectral windows covering
from 11.976 GHz to 18.224 GHz, each divided into 64 channels of 2 MHz
width.





In all the observations, 3c138 was used as primary flux calibrator,
while J0547+2721 was used as phase calibrator with 4 observations of
1min 30s between the target observations of 9 min. After looking at
the resulting images from each run to check the quality of the data,
we excluded one of the three observing runs with Ka band. 

\begin{figure*}[t]
\begin{center}
  \includegraphics[width=\linewidth]{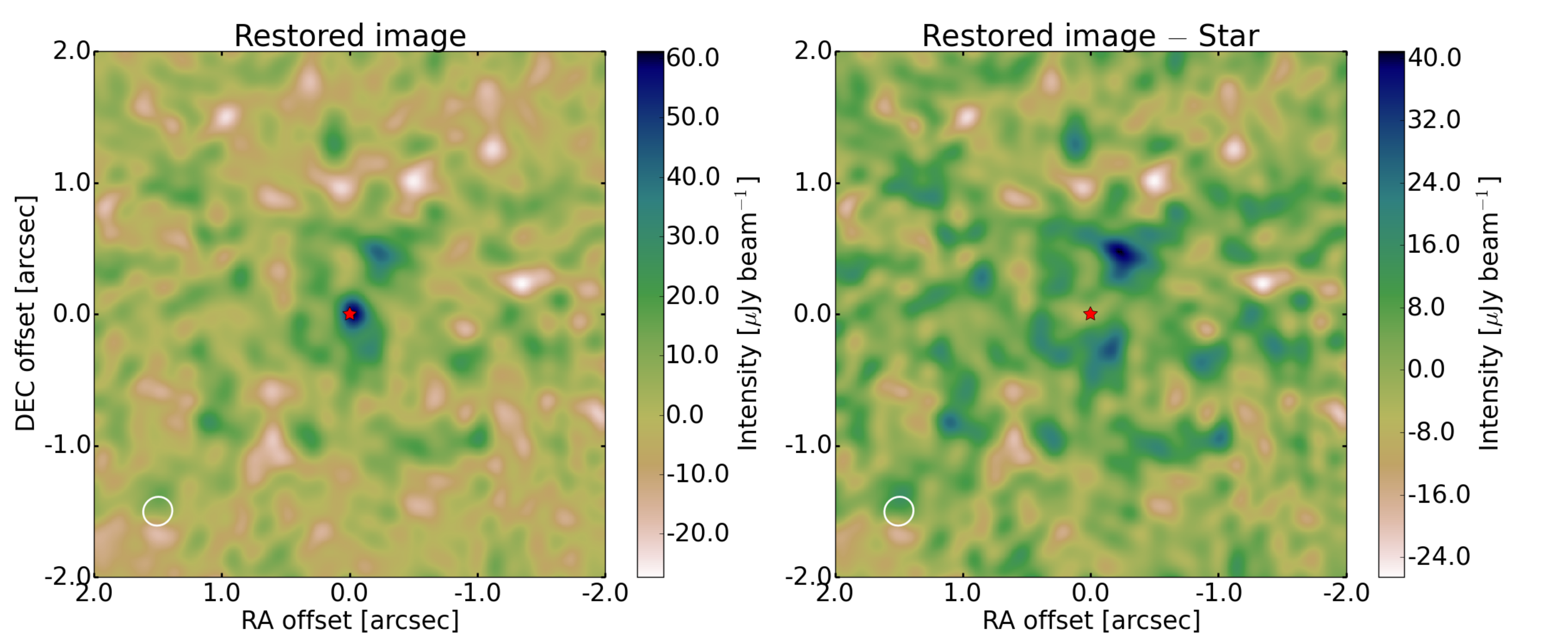}
  \caption{ \label{fig:VLA} Restored VLA Ka ($\sim$ 33 GHz)
    images. Left panel: restored image. Right panel: restored
      image after subtracting the stellar emission with a point source
      model to the visibilities. The beam size in the Ka images is
    $0.23''\times0.22''$ and is represented by a white ellipse in both
    images. The $x$ \& $y$ axes indicate the offset from the
      stellar position in RA and DEC in arcsec, i.e. north is up and
    west is right. The stellar position is marked with a red star.}
\end{center}
\end{figure*}

\subsubsection{VLA images}
In Figure \ref{fig:VLA} we present the restored VLA Ka image and
restored image after subtracting the central component with a point
source fit to the visibilities. The Ka image presents a noise level
($\sigma$) of 6.6 $\mu$Jy~beam$^{-1}$ and the beam size is
$0.23''\times0.22''$ with natural weights. Two bright compact
regions separated by $0.5''$ stand out from some extended emission,
which is not very well resolved due to low signal-to-noise. The
southern compact emission has a peak intensity of 63.0
$\mu$Jy~beam$^{-1}$ and is located at the position where the star is
expected according to the stellar coordinates and proper motion
\citep{Perryman1997}. The other compact emission is 44.0
$\mu$Jy~beam$^{-1}$ and is located to the north-west of the star,
where \citet{Isella2010} found an intensity maximum at millimeter
wavelengths. To the south of the star there is also disk emission with
a peak intensity of 30.0 $\mu$Jy~beam$^{-1}$, slightly higher than
4$\sigma$. This probably correspond to dust thermal emission from the
disk and puts an upper limit to the disk emission at this
frequency. However, the morphology of this emission is hard to
elucidate given the noise levels and resolution. The northern clump is
separated from the star by 0.5$''$, which at a distance of $\sim$280
pc translates to a de-projected radius of $\sim$150 au in the disk
plane. The total flux at 33 GHz inside a circumference of 1.0
  arcsec radius is $366\pm49 \mu$Jy and $289\pm33\mu$Jy after
  subtracting the stellar emission.

The Ku data only shows compact emission coming from the star location
with a total flux of 67.1$\pm7.7$ $\mu$Jy and an image noise level of
4.1 $\mu$Jy~beam$^{-1}$. The synthetic beam of this observation
corresponding to natural weights has a size of
$0.56''\times0.18''$. We are not showing the Ku image because it is
featureless and only shows a point source at the stellar position.

\subsubsection{Flux loss}

We neglect flux loss in the VLA data. To study this possibility in the
VLA Ka observations due to missing baseline spacing and limited
\textit{u-v} coverage, we simulate observations using the VLA Ka
\textit{u-v} coverage and a model image of the disk at 337 GHz (the
model is described in Section \ref{sec:model}). At this frequency the
disk emission is more extended than at 33 GHz, and thus flux loss is
more likely. We scaled the intensity levels of the model image at 337
GHz to have a simulated peak intensity equal to the Ka peak
intensity. Gaussian noise was added to the model visibilities to
reproduce the same noise level than the restored VLA Ka image. The
final simulated image has a total flux of 620$\pm27$ $\mu$Jy inside a
circumference of 1.0 arcsec radius, centered at the stellar
position. This is consistent with the scaled model flux of 609 $\mu$Jy
within the estimated errors. Therefore, we conclude that there is no
flux loss due to missing uv sampling in the case of the VLA data.




\subsection{ALMA}

To study the disk continuum emission at millimeter wavelengths we made
use of ALMA archival data in Band 7 of MWC~758 from the project
2011.0.00320.S \citep{Chapillon2015}. The observations were conducted
over two nights in August and October of 2012. The ALMA correlator was
set in Frequency Division Mode (FDM) to provide 4 spectral windows
divided in 384 channels, centered at 337.773 GHz with a total
bandwidth of 234.363 MHz, 344.469 GHz with a total bandwidth of
234.363 MHz, 332.488 GHZ with a total bandwidth of 233.143 MHz and at
330.565 GHz with a total bandwidth of 233.143 MHz. The observations
were carried with 29 antennas of 12 meter diameter. The minimum and
maximum projected baselines were 15 and 375 m and the total time on
source was 12 min. We flagged the lines CO $J=$3-2 (337.79599 GHz) and
$^{13}$CO $J=$3-2 (330.58797 GHz) to study the dust continuum
emission.

\begin{figure*}[t]
\begin{center}
  \includegraphics[width=\linewidth]{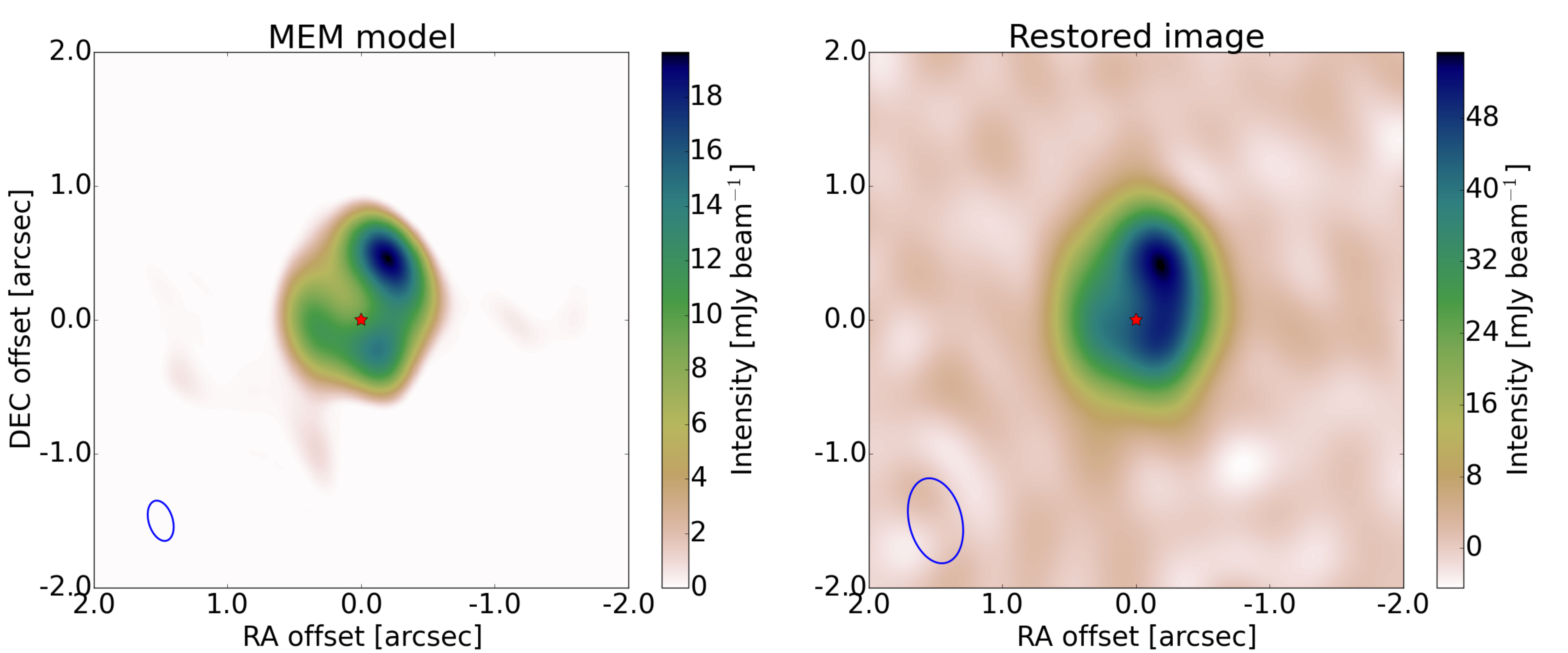}
  \caption{ \label{fig:ALMA} ALMA maps at 337 GHz (Band 7). Left
    panel: MEM non-parametric model (regularized with the maximum
    entropy method) with an approximate angular resolution of
    $0.31''\times0.18''$. Right panel: Restored image adding residuals
    and convolving with a Clean beam corresponding to \textit{briggs}
    weighting ($0.64''\times0.40''$). The respective beams are
    represented by blue ellipses. The $x$ \& $y$ axes indicate the
    offset from the stellar position in RA and DEC in arcsec,
    i.e. north is up and west is right. The stellar position is marked
    with a red star.}
\end{center}
\end{figure*}

In Figure \ref{fig:ALMA} we present the ALMA MEM image using uvmem and
a restored image that can be compared to a Clean image. The model
image has an approximate resolution of $0.31''\times0.18''$, derived
from the response to a point source. The restored image was produced
by convolving with a Clean beam of size $0.64''\times0.40''$
characteristic of \textit{briggs} weighting to achieve a better
resolution. In the restored image, the total flux is $210\pm5$
  mJy, with a peak intensity of 55.0 mJy~beam $^{-1}$ and a noise
  level of 1.3 mJy beam$^{-1}$.
  
In Table \ref{tab:disk_fluxes} we summarise the new flux values of
MWC~758 added by this work along with global spectral trends.

\begin{table}[h]
 \centering
   \caption{VLA and ALMA fluxes} 
   \begin{tabular}{ccc}
    \hline
    \hline
    Measurement &  Value  \\ 
    \hline
    ALMA band 7 flux (337 GHz) [mJy] & 205 $\pm$ 5  \\
    VLA Ka (33 GHz) flux [$\mu$Jy] & 366 $\pm$ 49 \\
    VLA Ka disk flux\footnote{Flux inside a circumference of 1.0 arcsec radius after subtracting the central component.} [$\mu$Jy] & 289 $\pm$ 33 \\ 
    230\footnote{Data from \citealt{Chapillon2008}.}-337 GHz spectral index & 3.4 $\pm$ 0.1 \\
    337-33 GHz spectral index & 2.8 $\pm$ 0.1 \\
    \hline
   \end{tabular} 
\label{tab:disk_fluxes}
\end{table}


\section{Analysis} \label{sec:Analysis}

The VLA Ka and ALMA maps show that the disk of MWC~758 is far from an
axisymmetric disk. The disk shows a peak of intensity to the
north-west side of the disk, centered at a distance of $0.5''$ from
the star, consistent with previous millimeter observations
\citep{Isella2010}. In both ALMA maps and more evident in the MEM
model, a local intensity minimum appears at the center of the disk,
confirming the presence of a cavity of millimeter grains. Unlike the
case of HD~142527 and IRS 48, the maximum and minimum of intensity are
not opposite in azimuth (azimuthal wavenumber m = 1)
\citep{Casassus2013Nat, Fukagawa2013, vanderMarel2013}. Moreover, a
second maximum in azimuth appears in the ALMA MEM model to
the south-west of the star, which translates to an azimuthal extension
to the south of the northern intensity maximum in the restored image.

\subsection{Spectral trends}

To investigate the origin of the emission in the VLA maps, it is
useful to compute the intra-band and inter-band spectral indexes,
defined as
$\alpha_{\nu_1}^{\nu_{2}}=\log(I_2/I_1)/\log(\nu_2/\nu_1)$. The Ka-Ku
inter-band spectral index ($\alpha_{15}^{33}$) of the emission coming
from the stellar location is $0.36\pm0.20$, while the Ku intra band
spectral index is $\alpha_{15}=0.8\pm0.6$, computed from a point
source fit to the Ku data at the different channels. Both values are
consistent with free-free emission from the central star associated
with a stellar wind or stellar accretion. Assuming a spectral index
between 0.5 and 1.0 from free-free emission and the VLA flux level at
33 GHz, we expect emission levels of 0.2-0.7 mJy~beam$^{-1}$ at 337
GHz, well below the disk emission and the noise level reached in the
ALMA data. Similar unresolved emission has been detected with ATCA
observation at the location of the central star in HD~142527 with a
slightly higher spectral index of 1.0$\pm$0.2
\citep{Casassus2015trap}.


On the other hand, the northern compact emission has an inter-band
spectral index of $\alpha_{Ka}^{B7}=3.1\pm0.1$ when we compare with
the intensity maximum of the restored ALMA image. Using
$\alpha_{Ka}^{B7}$, the predicted flux of the northern clump at 15 GHz
is $\sim$6.0 $\mu$Jy, slightly above the noise level of the Ku data.




Given the unknown temperatures, the uncertainties in the clump flux,
the non detection at 15 GHz, and the lack of spatial resolution in the
ALMA data, it is not possible to conclude on the spectral index
$\beta$ for an optical depth power law $\tau(\nu)\propto \nu^{\beta}$,
preventing us from studying a direct relation between the observations
and a dust size distribution in the northern clump.  Additional
observations with similar resolution levels are necessary to derive a
$\beta$ index to directly study the dust size distribution in the
clump. However, in Section \ref{sec:model} we implement a simple
parametric vortex model for the disk that reproduces the SED and
morphology of the VLA and ALMA maps. We use this model to interpret
the VLA clump in Section \ref{sec:model}.



\subsection{Comparison between ALMA and VLA Ka maps}
\begin{figure}[h!]
\begin{center}
  \includegraphics[width=\linewidth]{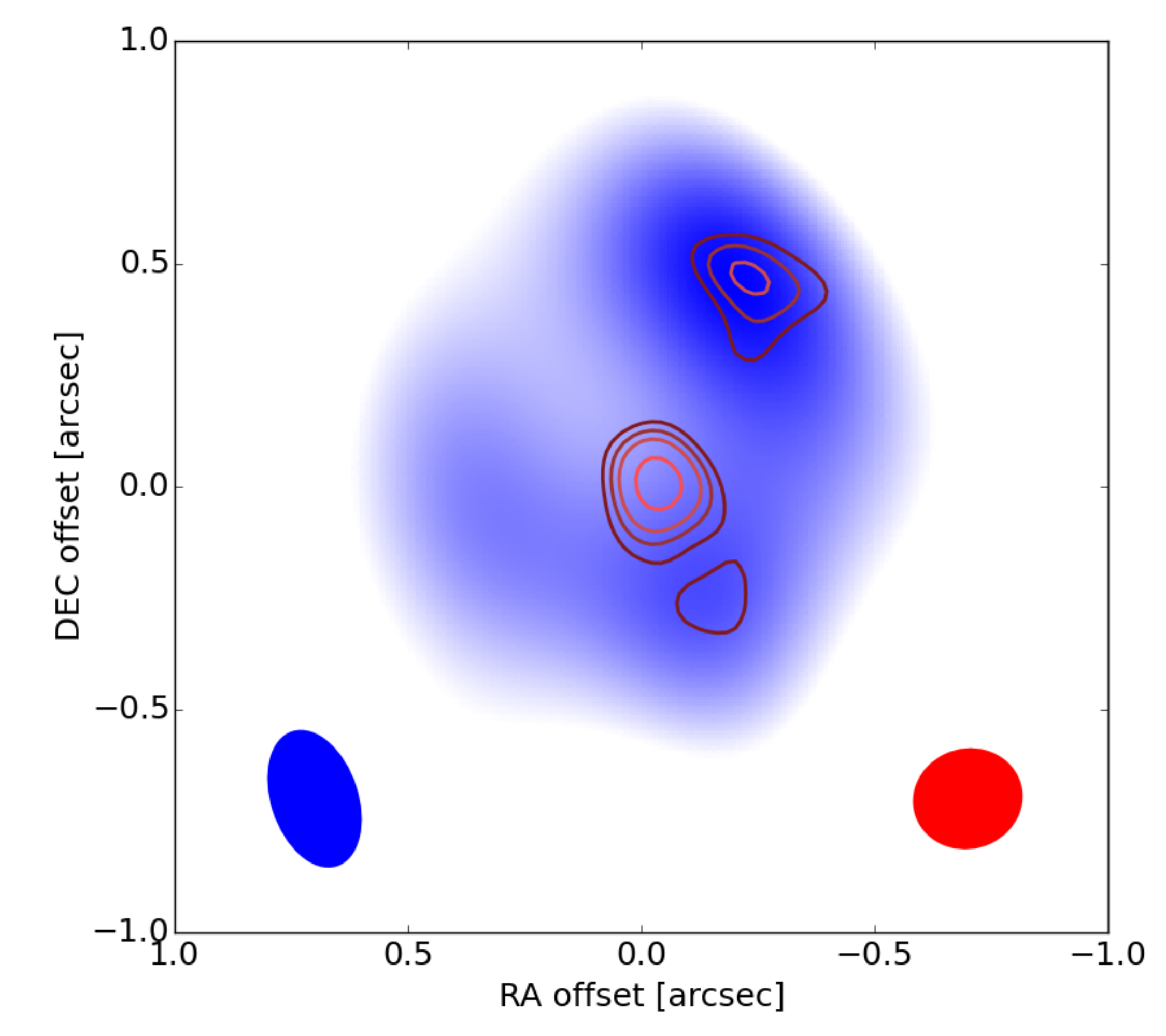}
  \caption{ \label{fig:ALMAVLA} ALMA-VLA overlay. In blue the ALMA
    band 7 MEM model in an arbitrary color scale, while in red the
    restored VLA Ka image contours set at arbitrary levels to
    highlight the morphology (the lowest contour level is 4.2
    $\sigma$). The beam size of the restored VLA VLA Ka image
      and ALMA MEM model is represented by a red and blue ellipse,
      respectively. The $x$ \& $y$ axes indicate the offset
    from the stellar position in RA and DEC in arcsec,
    i.e. north is up and west is right.}
\end{center}
\end{figure}

In Figure \ref{fig:ALMAVLA}, we present an overlay between the
restored VLA Ka image and the ALMA band 7 MEM model. The VLA Ka
northern peak intensity matches the location of the ALMA band 7
maximum. At 337 GHz we expect that most of the emission comes from
$\sim$millimeter-sized grains. If this intensity maximum is tracing a
maximum in the dust density distribution of millimeter-sized grains,
the VLA and ALMA maximum could be due to dust grains being
trapped in a pressure maximum of the gas. In the dust trap scenario,
we would expect to observe at centimeter wavelengths a higher contrast
between the intensity maximum and the rest of the disk and a more
compact emission, probing the distribution of centimeter-sized grains
that are more easily trapped in gas pressure maxima. However, the ALMA
maps are not as well spatially resolved as the VLA Ka image due to
differences in the \textit{u-v} coverage. In
Sec. \ref{sec:almavla_samer} we compare both datasets at the same
resolution level.

Figure \ref{fig:ALMAVLA} also shows disk emission to the south of the
stellar position in the VLA Ka map that matches a second peak
intensity in the ALMA MEM model. This peak is closer in than the
northern clump, at an angular distance of $\sim0.3''$. However this
local maximum disappears in the restored ALMA image due to the larger
 beam.

\subsection{ALMA and VLA Ka map at the same resolution level} \label{sec:almavla_samer}
\begin{figure}[h!]
\begin{center}
  \includegraphics[width=\linewidth]{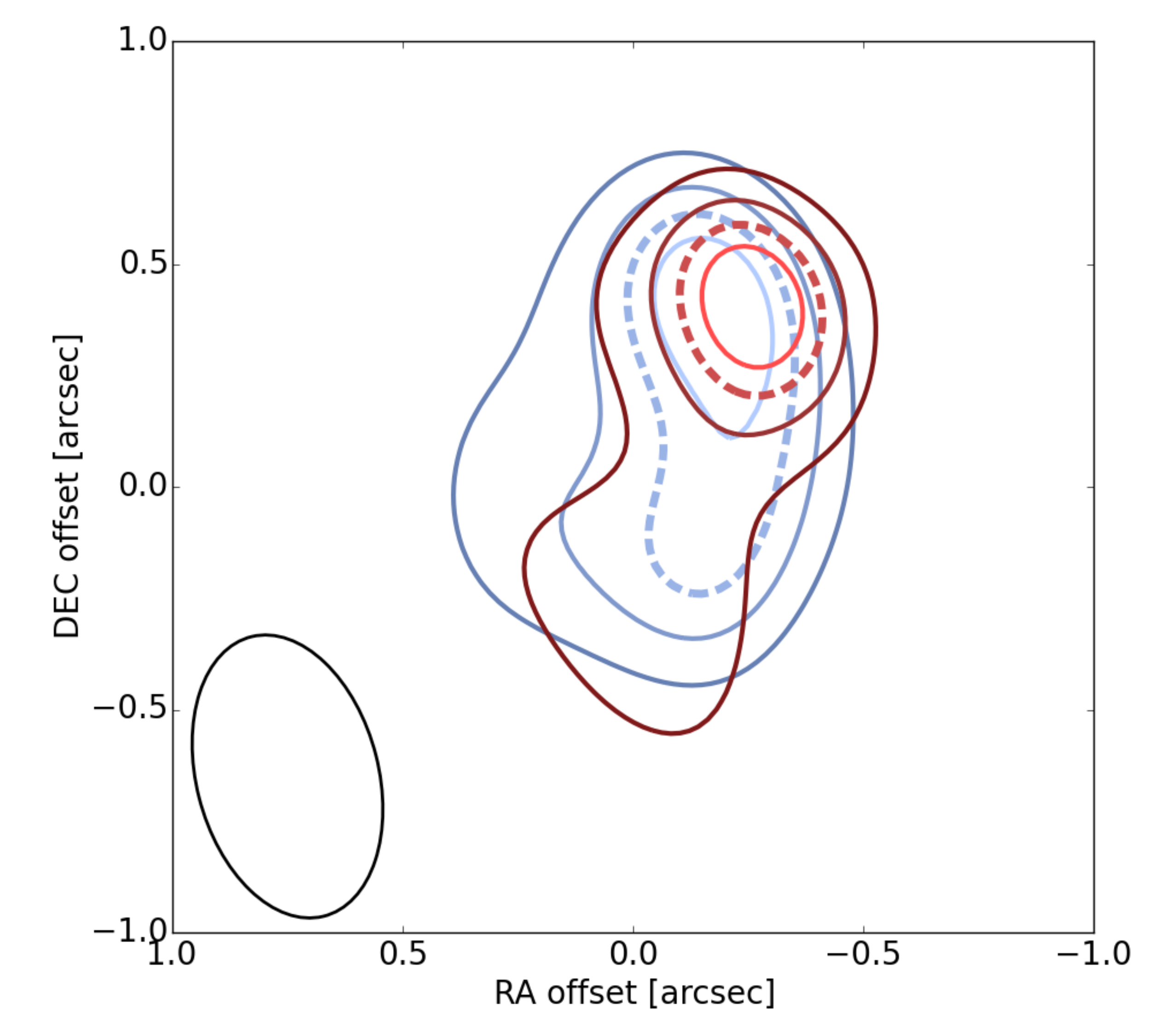}
  \caption{ \label{fig:ALMAVLA_con} ALMA-VLA map contours
    overlay. Blue contours: ALMA Band 7 restored image. Red contours:
    degraded VLA image (restored image after subtracting the star and
    convolving with the ALMA beam). The contour levels are 0.6, 0.75,
    0.85 and 0.92 times the peak intensity of each map. The lowest
    contour level of the VLA map represents $\sim$3$\sigma$ at this
    resolution. The dashed red and blue lines represent the
      0.85 peak intensity level. The beam size of both maps is
      represented by a black ellipse. The $x$ \& $y$ axes indicate
    the offset from the stellar position in RA and DEC in
    arcsec, i.e. north is up and west is right. }
\end{center}
\end{figure}

To bring both datasets to comparable resolutions, we convolved the
restored VLA Ka image with the ALMA Clean synthetic beam of size
$0.64''\times0.40''$ after subtracting the stellar
  emission. We call this map the degraded VLA image. In Figure
\ref{fig:ALMAVLA_con} we show the contour levels 0.6, 0.75, 0.85 and
0.92 peak intensity of the degraded VLA and restored ALMA images. The
degraded VLA image presents a morphology similar to the ALMA MEM model
with two peak intensities: to the north-west and south of the star
(see Fig. \ref{fig:ALMA}). The northern clump in the degraded VLA
image is still more compact than in the restored ALMA image, with a
larger contrast. The solid angle inside the 0.85 intensity maximum
contour in the degraded VLA image is 0.09 arcsec$^2$, while it is 0.23
arcsec$^2$ in the ALMA map. This result supports the dust trap
  interpretation, finding that the 33 GHz dust emission is more
  compact than at 337 GHz.

The peak intensity to the south of the star in the restored VLA Ka
image remains after star subtraction and convolution with the ALMA
beam, with a morphology similar to that observed in the ALMA MEM
model. The nature of this compact emission is not clear as it is just
$\sim$4 times the noise level in the restored VLA Ka image. At the
ALMA resolution the cavity is hard to distinguish, although the peak
intensity is offset from the stellar position in both maps, suggesting
a cavity depleted of big grains.

To confirm the dust trap interpretation, ALMA observations in extended
configuration are required to resolve the disk with a similar or finer
resolution than the VLA Ka observations. Deeper VLA observations are
also necessary to detect the rest of the disk.

\subsection{Parametric non-axisymmetric model} \label{sec:model}

We develop a parametric non-axisymmetric disk model, inspired by the
steady state vortex solution to the gas and dust distribution
described in \cite{Lyra2013} (hereafter L$^2$13). The aim is to
reproduce part of the morphology seen in the ALMA and VLA maps, and
the global SED. The model consists of a central star surrounded by a
disk of gas and dust. We model the star using a Kurucz template
spectrum \citep{2003IAUS..210P.A20C} with a temperature of 8250 K and
a stellar radius of 2.6 $R_{\odot}$, to fit the SED in the optical
with a visual extinction of 0.22 \citep{vandenAncker1998}. We define
two distinct disk zones (see Table \ref{tab:disk}):

\begin{enumerate}
  \item A dusty inner disk with small grain sizes ranging from
    0.1 to 1 $\mu$m, composed of amorphous carbon and silicate grains
    to reproduce the NIR excess.
  \item A non-axisymmetric outer disk composed of amorphous carbon and
    silicate grains to reproduce part of the disk morphology seen in
    the VLA and ALMA images, and the photometry from the SED in the
    FIR and radio.
\end{enumerate}

\begin{table}[h]
 \centering
   \caption{Disk parameters} 
   \begin{tabular}{ccc}
    \hline
    \hline
    Parameter &  Inner disk   & Outer Disk\\ 
    \hline
    $r_{\min}$ [au]      &  0.2  & 20.0  \\
    $r_{\max}$ [au]      &  80.0 & 300.0  \\
    $\Sigma(r)$         &   $\Sigma_c\left(\frac{r}{r_c}\right)^{-0.5}$   & Eq. \ref{eq:doutdisk}   \\
    $H(r)$ [au]         &   0.2$\left(\frac{r}{1.0 \ \mathrm{au}}\right)^{1.1}$  & $19.0\left(\frac{r}{100.0 \ \mathrm{au}}\right)$ \\
    $M_{d}$\footnote{The dust size distribution is $n(a)\propto a^{-3.5}$. The dust internal density for silicate and amorphous carbon grains is 4.0 and 2.0 g/cm$^3$ respectively.} [$M_{\odot}$]             &   $5.0\times10^{-8}$  & 7.1$\times10^{4}$ \\
    Mass fraction in Silicates &  0.9  & 0.48  \\
    $a_{min}$ [$\mu$]    & 0.1   & 1.0   \\
    $a_{max}$ [$\mu$]    & 1.0  & 4$\times10^{3}$  \\
    \hline
   \end{tabular} 
\label{tab:disk}
\end{table}

The total gas mass in the disk is 0.1 $M_{\ast}$. The outer disk
incorporates an analytic prescription for a steady state vortex
defined for both gas and dust distribution, based on the work of
L$^2$13. We first define an axisymmetric gas background inspired by the
surface density distribution used in \citet{Isella2010}, and following
the solution for the surface density of a viscous Keplerian disk
\citep{Lynden-Bell1974}
\begin{small} \begin{eqnarray}
  \Sigma_{b}(r)=\Sigma_{c} \left(\frac{r}{r_{c}} \right) ^{-\gamma}  \exp\left( -\frac{1}{2(2-\gamma)} \left[ \left(\frac{r}{r_{c}} \right)^{2-\gamma} -1 \right] \right), \\
  \rho_{b}(r,z,S)=\frac{\Sigma_{b}(r)\sqrt{S+1}}{\sqrt{2\pi}H}\exp\left[-\frac{z^2}{2H^2}(S+1)\right],
\end{eqnarray} \end{small}

\noindent where $H$ is the scale height of the disk and $S$ is defined
as the ratio between the the Stokes number (St) and $\delta_t$, a
turbulent diffusion parameter that measures the strength of the
turbulence in the disk. We assume a global $\delta_t$ in the disk for
simplicity, although the turbulence in the core of the vortex can be
driven by different mechanisms than in the rest of the disk. We set
$r_c=120.0$ au and $\gamma=-2.0$ to form an outer disk that
matches the peak intensity and cavity in the ALMA observations, and
the SED at long wavelengths, similar to the profile derived in
\citet{Isella2010}. However, these parameters are not well constrained
due to the lack of spatial resolution in the ALMA data and SED
degeneracies.

We add the vortex steady state solution described by eq. 64 in L$^2$13
with a slight modification to include a global disk gas density
field. We impose that the gas density described by the vortex solution
has to match the gas background at the vortex boundary where the
vortex flow becomes supersonic with respect to the disk. This happens
when $a$ (defined in L$^2$13 as the axial elliptical coordinate
corresponding to the vortex's semi-minor axis) is equal to
$\frac{H}{\chi \omega_V}$, where $\chi$ is the vortex aspect ratio and
$\omega_V=\Omega_V/\Omega$ is the dimensionless vortex frequency. The
same applies to the dust, and a general equation for both gas and dust
can be written as \begin{small}
\begin{eqnarray}
    &\rho(r,\phi,z)& = \rho_{b}(r,z) \max\left\{ 1, \  c \ \rho_{V}(r,\phi) \right\} \label{eq:diskdensity}, \\
    &\rho_V&=\exp\left[- \left( \frac{ (r-r_0)^2}{2H_v^2}+\frac{r_0^2\phi^2}{2H_v^2 \chi^2} \right) \right], \\
    &c&= \exp\left[ \frac{f^2(\chi)}{2\chi^2\omega_v}(S+1) \right], 
\end{eqnarray} \end{small}

\noindent where $c$ is the contrast between the density maximum inside
the vortex and the background density field $\rho_{b}$ at the same
radius (see Section \ref{appendix} for more details). $c$ is defined to have a
global density field that matches the background density field at the
vortex boundary. $H_v=H/(f(\chi) \sqrt{S+1})$ is the vortex scale
length, where $f(\chi)$ is a scale function defined in L$^2$13 and is a
function of $\chi$. A smooth solution for the density field,
considering a smooth shock perimeter can be written as
\begin{equation} \label{eq:doutdisk}
    \rho(r,\phi,z,S) = \rho_{b}(r,z)[ 1 + (c-1) \rho_{V}(r,\phi,z,S)].
\end{equation} 

We set $\chi=5$ (which is not well constrain, but between the range of
  accepted values to remove the elliptical instability at the
  vortex centre, \citealt{Lesur2009}) and $\delta_t=3\times10^{-3}$. To
compute $\omega_V$ we use the Goodman-Narayan-Goldreich solution
\citep{Goodman1987}. 

\begin{figure}[!t]
\begin{center}
  \includegraphics[width=\linewidth]{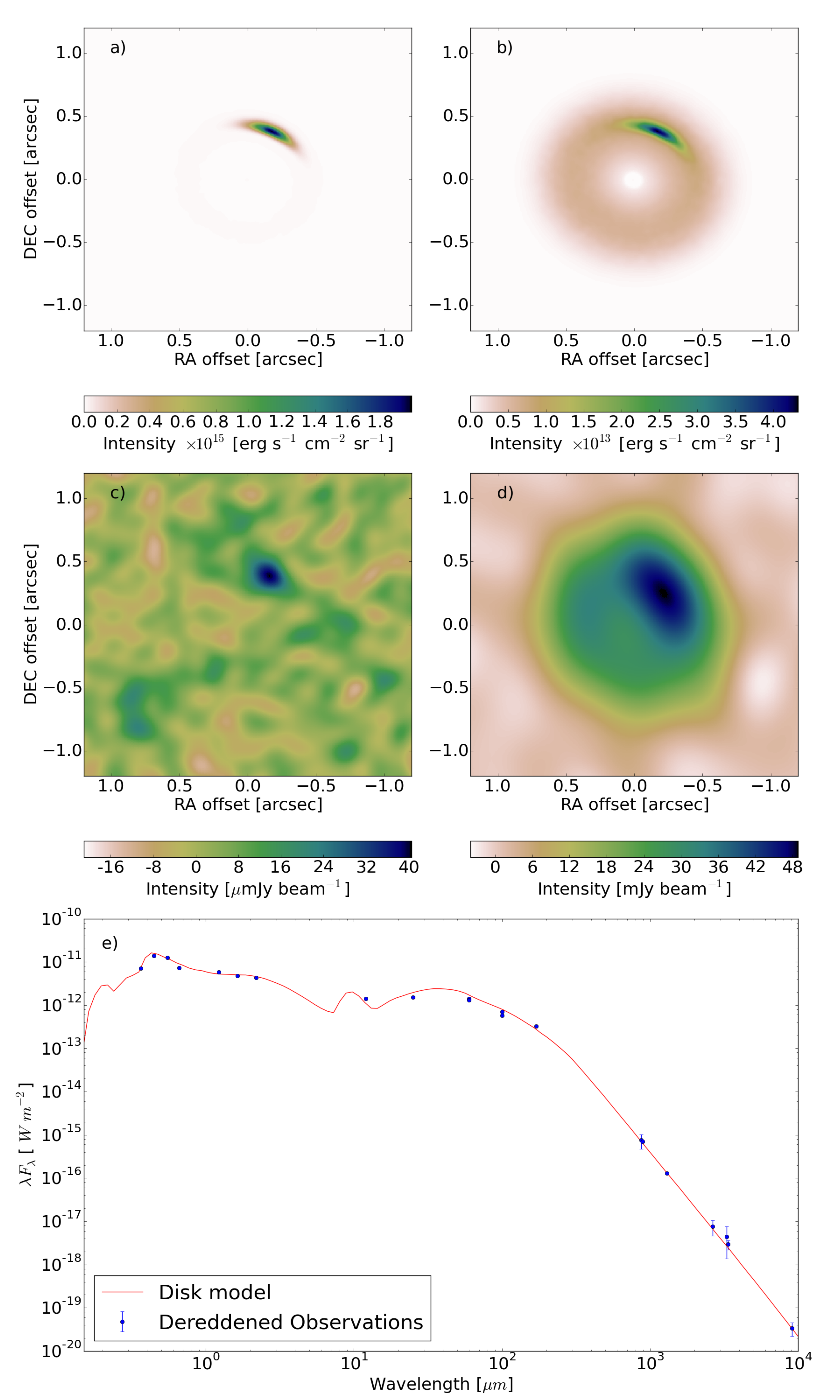}
  \caption{ \label{fig:RADMC} Parametric model predictions. (a) 33 GHz
    model image. (b) 337 GHz model image. (c) Simulated VLA Ka
    observation using the model image. (d) Simulated ALMA band 7
    observation using the model image. (e) Dereddened SED of MWC~758
    (blue dots) compared with the model (red line). Photometric data
    points were taken from \citet{Hog2000, Cutri2003, Elia2005,
      Chapillon2008, Isella2010} and the new VLA Ka and archival ALMA
    band 7 data presented in this work. We deredden the observations
    assuming a visual extinction $A_V$=0.22 \citep{vandenAncker1998}
    and an total-to-selective extinction ratio of $R_V$=3.1. The
    errorbars represent 2$\sigma$ errors. In (a)-(d) the $x$ \&
    $y$ axes indicate the offset from the stellar position in
    RA and DEC in arcsec, i.e. north is up and west is right.}
\end{center}
\end{figure}

Using Eq. \ref{eq:diskdensity} and reasonable values for $\delta_t=
10^{-2}-10^{-3}$, it is impossible to produce a vortex with
  the contrast required to reproduce the asymmetry seen in the ALMA
  maps. To increase the contrast with the disk background, we
  artificially extend the vortex steady state solution beyond its
  original boundary to obtain a better match with the observations
(see Sec. \ref{sec:discussion} for a discussion on this). This
translates into changing $c$ to $\exp\left[
  r_s^2\frac{f^2(\chi)}{2\chi^2\omega_v}(S+1) \right]$, where $r_s$ is
the ratio between the increased vortex size and the vortex original
boundary (see Appendix for a detailed description). We set
  $r_s=3$ to have a good match between the restored ALMA image and
the model image convolved with the synthetic ALMA beam (see Figures
\ref{fig:ALMA} and \ref{fig:RADMC}d).


We implement the model using 9 dust species in total: 2 dust species
for the inner disk, representing silicates and amorphous carbon
grains; 7 dust species for the outer disk, one accounting for small
amorphous carbon grains (from 1 to 10 $\mu$m), while the rest
represent silicate grains from 1 $\mu$m to 4 mm with different spatial
distributions set by eq. \ref{eq:doutdisk}. The opacities for each
dust species in the model are computed using the 'Mie Theory' code
written by \citet{BohreHuffman1983} considering a dust size
distribution with a power law index of $-3.5$, and optical constants
for amorphous carbon \citep{Li_Greenberg_1997A&A...323..566L} and
``Astronomical silicate'' \citep{Draine2003}. The emergent intensities
from our model at different frequencies were computed using RADMC-3D
version
0.39\footnote{http://www.ita.uni-heidelberg.de/∼dullemond/software/radmc-3d/}
\citep{RADMC3D0.39}, using the \citet{Henyey1941} parametrization of
the phase function of scattering.

In Figure \ref{fig:RADMC} we present the model images at 33 and 337
GHz at the top. The middle images show the simulated observations of
the model images with the same \textit{u-v} coverage and noise levels
of the VLA Ka and ALMA observations. In addition, we present the model
SED that matches roughly the observations. As expected from the vortex
model, at long wavelengths the disk emission is concentrated in the
pressure maximum as it traces the distributions of bigger grains which
are highly trapped inside the vortex.

\begin{figure}[!t]
\begin{center}
  \includegraphics[width=\linewidth]{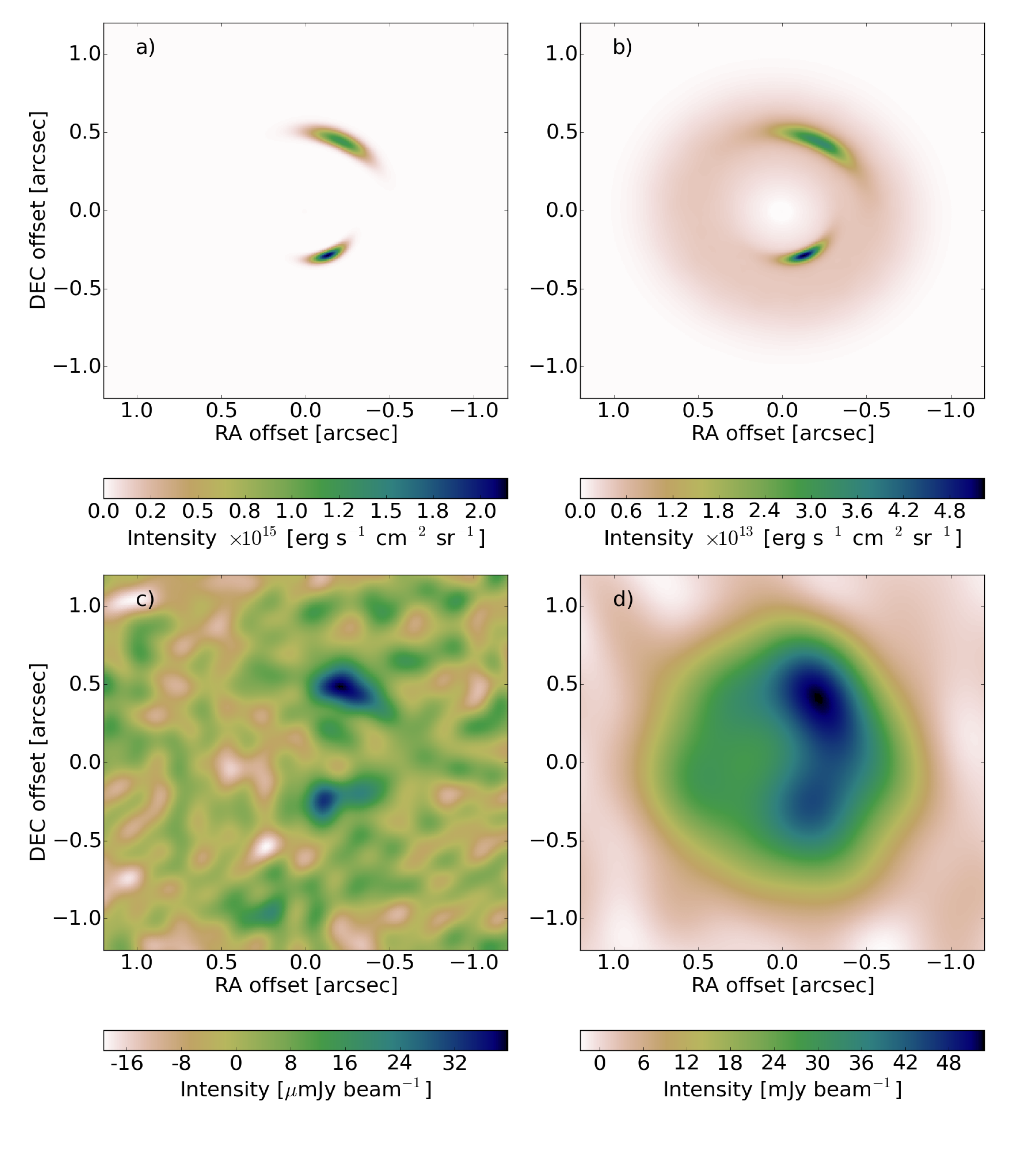} 
  \caption{ \label{fig:RADMC_2v} Parametric model predictions
      with two vortices. (a) 33 GHz model image. (b) 337 GHz model
      image. (c) Simulated VLA Ka observation using the model
      image. (d) Simulated ALMA band 7 observation using the model
      image. In (a)-(d) the $x$ \& $y$ axes indicate the offset
      from the stellar position in RA and DEC in arcsec, i.e. north
    is up and west is right.}

\end{center}
\end{figure}

The simulated ALMA image shows a peak of intensity similar to the
restored ALMA image, although the morphology is not exactly the
same.  An interesting difference is the second intensity
maximum seen in the ALMA MEM image, which appears as an extension to
the south of the maximum in the restored image. This cannot be
  reproduced by a single and unresolved density maximum as our model
  shows. A second vortex would be needed to account for this
emission. In Figure \ref{fig:RADMC_2v} we present the model
  predictions and simulated observations of a model with two vortices
located at the position of the VLA northern and southern clumps. The
southern extension of the ALMA peak intensity is naturally reproduced,
and also a decrement appears at the north-east side of the disk due to
the elongated beam (see figure in Sec. \ref{app:MEM} in
  Appendix). The VLA simulated observations show two unresolved
clumps at the location of the VLA northern and southern
clumps. The disk parameters are the same to the ones described
  above. The second vortex is located at a radius of 90 au and it has
  a size ratio $r_s$ of 3.12.


If the northern clump emission seen in MWC~758 in band 7 and Ka data
is produced by a dust trap, it is a possible location for planetesimal
and planetary core formation, as grains can grow faster and avoid the
inward drift. From our model the dust temperature at the clump
position is $\sim$30 K, but in the observations the brightness
temperature of the clump at 33 GHz is 1.0 K. Thus it is highly
unresolved or it is optically thin at this frequency. From our
synthetic model, the inferred dust mass inside the vortex boundary is
$\sim45$ $M_{\earth}$, enough mass to form several planetesimals that
could lead to the formation of a planetary core of a gas
giant. However, grain porosity is not included in our model, thus this
inferred mass is probably higher than the real value.


\subsection{Comparison with SPHERE PDI data}\label{sec:Sph}

We also compare the VLA and ALMA observations with publicly available
VLT SPHERE/IRDIS PDI (polarimetric differential imaging) data
\citep{Benisty2015}. The data was taken on December 5, 2014 during the
science verification time of SPHERE. MWC 758 was observed in DPI (dual
polarimetric imaging) mode in \textit{Y}-band (1.04 $\mu$m) with the
IRDIS sub-instrument of SPHERE, with a 185 mas diameter coronagraphic
mask and a Lyot apodizer. 70 exposures of 64 seconds each were taken,
of which 48 (total integration time 51.2 minutes) were used for the
reduction presented in this paper. PDI has proven to be a powerful
technique to suppress the stellar light, and to first order
approximates the reflected light off the disk surface by only
measuring the polarized fraction of the light
\citep{Avenhaus2014}. The frames were dark-subtracted, flat-fielded
and bad pixels were removed, then centered. The polarized flux was
calculated and instrumental effects accounted for in a way similar to
that described in the Appendix of \citet{Avenhaus2014}. In Figure
\ref{fig:SPHALMAVLA}, the resulting SPHERE image reveals the disk
outside of the inner working angle set by the coronagraph, including a
strong, two-armed spiral.

\begin{figure}[h]
\begin{center}
  \includegraphics[width=\linewidth]{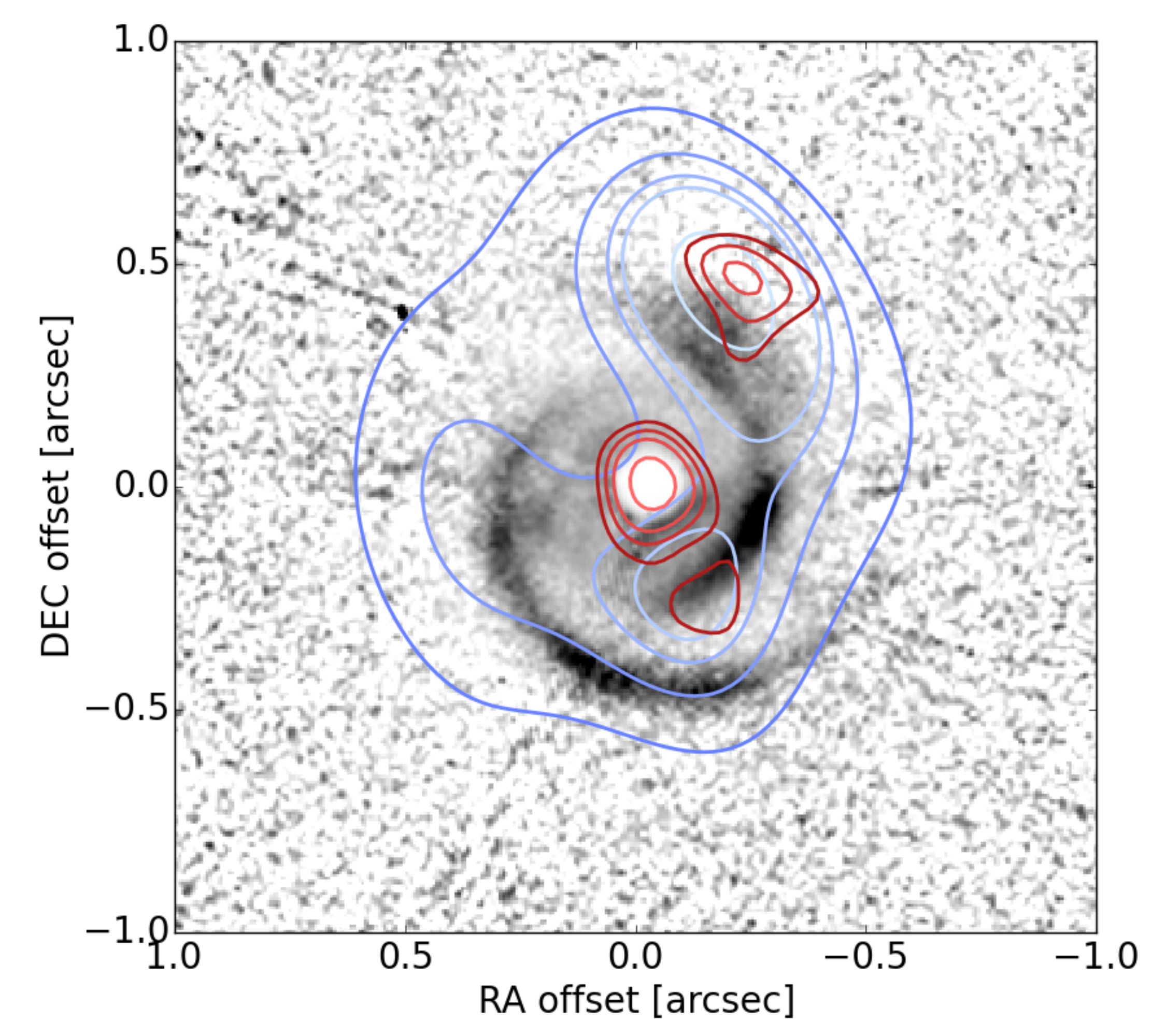}
  \caption{ \label{fig:SPHALMAVLA} SPHERE/IRDIS-ALMA-VLA
      overlay. In grey scale, the polarised intensity scaled by
      $r^{2}$ to highlight the surface density of the disk, obtained
      with SPHERE/IRDIS PDI. The decrement in the center correspond
    to the position of the coronagraph. The ALMA band 7 MEM map is
    represented in blue contours, while the restored VLA Ka map is
    presented in red contours. The contour levels are arbitrary to
    emphasise the disk morphology. The $x$ \& $y$ axes indicate the
    offset from the stellar position in the RA and DEC in
    arcsec, i.e. north is up and west is right.}
\end{center}
\end{figure}

Unlike the case of HD~142527, where a large cavity of $\sim$130 au is
seen in scattered light \citep{Avenhaus2014} and in the sub-mm
\citep{Casassus2013Nat, Fukagawa2013}, exposing the inner rim of the
outer disk to stellar radiation, the SPHERE PDI \textit{Y}-band image
of MWC~758 shows that, down to the inner working angle (26 au), the
cavity is not depleted of $\mu$m-sized dust grains traced at this
wavelength.  Moreover, the polarized surface brightness drops abruptly
behind the spiral arms, as if the arms shadowed the outer disk
\citep{Grady2013}. The presence of $\mu$m size dust grains could be
linked to flaring of the gas disk, the mass of a possible gap-opening
planet \citep{Fouchet2010}, or we might be looking at different stages
of similar systems.


In the same figure, contours from the ALMA band 7 MEM and restored VLA
Ka image are overplotted in blue and red, respectively. No evident
correlation appears between the most southern spiral arm and the dust
emission at 337 and 33 GHz, although we lack spatial resolution in the
ALMA data to trace fine structures. On the other hand, the disk
emission to the south of the star in the ALMA and VLA maps matches
part of the western arm, while the VLA northern clump maximum is
slightly offset ($\sim0.2''$) from the SPHERE PDI maximum at the same
PA. Such a radial displacement could be explained if there were a
large wall of dust with a higher scale height shadowing the disk at
the clump radius. This is probably related to a moderate or low
flaring in the outer disk.

The SPHERE image also shows a region devoid of scattered light
  emission to the north-east between two bright arms, matching also a
decrement in the ALMA MEM model that extends from the outer disk to
the stellar position. The origin of this decrement could be related
to a shadowing effect rather than a lack of material.


\section{Discussion} \label{sec:discussion}



MWC~758 shows spiral arms, non-axisymmetric dust thermal emission and
asymmetries in the CO emission. The origin of all the disk asymmetries
is hard to elucidate. A possible explanation for the non-axisymmetric
disk seen at millimeter wavelength is a dust trap caused by a
pressure maximum. A steady state vortex as the one proposed in L$^2$13
cannot produce such a high contrast between the dust density in the
vortex core and the disk background. A solution could be linked to the
vortex tail which also causes dust trapping and can extend farther out
from the vortex core \citep{Lesur2009, Lesur2010,
  Railton2014}. Recently, \cite{Surville2015} showed that
  vortices larger than the scale height might exist. Moreover,
numerical MHD simulations have shown that high contrasts in the gas
density field can be reached when large vortices form at the edge of
planet induced gaps, surviving for $\sim$1000 orbits
\citep{Zhu_Stone_2014ApJ...795...53Z}.

On the other hand, the azimuthal dust density structure in the outer
disk is probably more complex than an azimuthal wavenumber m = 1. The
second intensity maximum in the ALMA MEM model could be related to a
second vortex at the inner edge of the outer disk. MHD simulations
predict multiple vortices to form by the Rossby Wave Instability at
the inner edges of ``dead zones'' \cite{Lyra2012}. Moreover, the
decrement to the north-east in the ALMA and SPHERE images cannot be
easily explained by a decrement in the dust density and a single
density maximum.

An alternative explanation to the asymmetry in the dust continuum
emission at 337 GHz is that there is a strong asymmetry in the dust
temperature field in the outer disk, similar to HD~142527 where a
large crescent shows two peak intensities separated by a decrement in
temperature caused by shadows cast by a tilted inner disk
\citep{Casassus2015trap, Marino2015a}. If the possible
  decrement in MWC~758 is produced by a tilted or warped inner disk we
  would expect to see clear decrements almost opposite in azimuth,
  similar to the decrements in HD~142527 seen in polarized light
  \citep{Avenhaus2014}. Another possibility is that the spiral arms
seen in scattered light \citep{Grady2013, Benisty2015} cast shadows in
the outer disk at different azimuth and radii. Moreover, a combination
of both scenarios could be present. This could explain the asymmetric
CO emission \citep{Isella2010} as a temperature effect, the formation
of spiral arms due to tidal forces from an inclined inner disk, or
produced as a dynamic consequence of a perturbation of the gas
temperature field \citep{Montesinos2015}. However, the compact
emission to the north-west of the star at centimeter wavelengths seen
in the VLA maps suggests that the dust concentration is real, although
azimuthal structure in the dust temperature field is expected as the
disk is not axisymmetric.

The compact emission in the VLA Ka data could also be explained by a
companion object, e.g. a forming planet accreting
material. This planet could be also the responsible for the
  observed spiral arms in scattered light, as it is close to the
  planet proposed by \cite{Dong2015}. Such forming planet would have
higher temperatures than a passive disk as it is accreting, also
heating the disk around it. ALMA multi-band continuum and gas line
observations would be ideal to disentangle if the VLA and ALMA band 7
asymmetries are density or temperature effects and if the VLA clump is
accreting or not. Additional H-$\alpha$ high contrast imaging could
reveal the presence of an accreting companion.

\section{Conclusions} \label{sec:Conclusions}

We report new VLA observations at 33 and 15 GHz observations of the
disk around the star MWC~758. The disk emission at these wavelengths
traces the distribution of centimeter-sized dust grains. We found a
compact disk emission at the same location where previous SMA and
CARMA observations found deviations from an axisymmetric model at
shorter wavelengths \citep{Isella2010}. This compact emission is at
$0.5''$ ($\sim150$ au) from the central star.

The VLA data also shows a strong compact emission coming from the star
position, characterised by a spectral index of $\sim$0.5. This is
likely due to free-free emission from the central star associated
with a stellar wind or stellar accretion.

We compare these data with ALMA archival data, producing ALMA band 7
continuum deconvolved and restored images with a finer resolution than
previous SMA and CARMA observations. Both maps show non-axisymmetric
disk emission, with a peak intensity at the same location as the VLA
clump, but more extended. These multiwavelength observations fit in
the context of dust trapping, as the bigger grains traced at lower
frequencies are more concentrated around pressure maxima than the ALMA
emission around the maximum. However, in the ALMA maps the disk is not
as resolved as in the VLA data. New ALMA long-baseline observations
are needed to resolve better the disk morphology at millimeter
wavelengths and to confirm the dust trapping effect. 

In addition, we compare these observations with a new reduction of
archival VLT SPHERE PDI images. There is an offset of $\sim$0.2$''$
between the VLA clump and the north-western spiral feature. The same
arm matches a second intensity maximum in the dust continuum seen in
the VLA and ALMA maps. More interestingly, the decrement to the
north-east in the ALMA MEM model matches a decrement in the polarized
intensity, suggesting a shadowing effect rather than lack of material.

We develop a parametric non-axisymmetric model for the disk,
incorporating a steady state vortex solution based on L$^2$13. Such a
vortex model cannot explain the decrement seen to the
north-east of the star and a second maximum seen in the VLA and ALMA
data. To reproduce the peak intensity of the ALMA image, we required
to extend the vortex solution artificially beyond the shock
perimeter. We also find that a second vortex could explain the
extension to the south of the ALMA peak intensity and the decrement to
the north-east as an image reconstruction artefact.

MWC~758 is a very complex protoplanetary disk and there are still
details to be addressed: what is the link between the possible vortex
or vortices, and the spiral arms and apparent decrements? The missing
piece of the puzzle might be found with a detailed study of the disk
kinematics.



\acknowledgments

SM, SC, SP and HA acknowledge support from the Millennium Science
Initiative (Chilean Ministry of Economy), through grant ``Nucleus
P10-022-F''. SM acknowledges CONICYT-PCHA/MagísterNacional/2014 -
folio 22140628.  SC, SP and HA acknowledge financial support from
FONDECYT grants 1130949, 3140601 and 3150643, respectively. PER thanks
to the Chilean Postdoctoral Fondecyt project number 3140634 and
ALMA-Conicyt project number 31120006. CMW acknowledges support from
ARC Future Fellowship FT100100495. STM acknowledges the support of the
visiting professorship scheme from Universit\'e Claude Bernard Lyon 1.

This paper makes use of the following ALMA data:
ADS/JAO.ALMA\#2011.0.00320.S. ALMA is a partnership of ESO
(representing its member states), NSF (USA) and NINS (Japan), together
with NRC (Canada) and NSC and ASIAA (Taiwan) and KASI (Republic of
Korea), in cooperation with the Republic of Chile. The Joint ALMA
Observatory is operated by ESO, AUI/NRAO and NAOJ.

\begin{appendix}
  
  \section{Vortex model} \label{appendix}
  Following the procedure of L$^2$13, to match the the vortex steady
  state solution from L$^2$13 with a gas density background, we impose
  both density profiles to match at the vortex boundary where the
  vortex flow becomes supersonic. This happens when $a$ (defined in
  L$^2$13 as the axial elliptical coordinate corresponding to the
  vortex's semi-minor axis) is equal to $\frac{H}{\chi \omega_V}$,
  where $\chi$ is the vortex aspect ratio and
  $\omega_V=\Omega_V/\Omega$ is the dimensionless vortex
  frequency. Assuming the vertical dependence is the same inside and
  outside the vortex, we can focus only in the midplane. The density
  field inside the vortex can be written as
  \begin{equation}
    \rho(a,S)=c \rho_b  \exp\left[ - \frac{a^2 f^2(\chi)}{2H^2}(S+1)\right],
  \end{equation}
  where $c$ is a constant that has to be determined imposing the
  following condition
  \begin{eqnarray}
    \rho_{b}&=& \rho(a=\frac{H}{\chi \omega_V},S), \\
    \rho_{b}&=&c \rho_b \exp\left[-\frac{f^2(\chi)}{\chi^2 \omega_V^2}(S+1) \right],
  \end{eqnarray}
  finally obtaining $c=\exp\left[\frac{f^2(\chi)}{\chi^2
      \omega_V^2}(S+1) \right] $. Thus, the global solution is
  \begin{equation}
    \rho(r,\phi,z) = \rho_{b}(r,z) \max\left\{ 1, \  c \ \exp\left[ - \frac{a^2 f^2(\chi)}{2H^2}(S+1)\right] \right\}.
  \end{equation}

  A more realistic or smoother solution can be written as
  \begin{equation}
    \rho(r,\phi,z) = \rho_{b}(r,z) \left( 1+ \  (c-1) \ \exp\left[ - \frac{a^2 f^2(\chi)}{2H^2}(S+1)\right] \right) \label{eq:Gsol}.
  \end{equation}

  If we extend the vortex from its original boundary to $\frac{r_s
    H}{\chi \omega_V}$, $c$ has to be redefined as
  $c=\exp\left[\frac{r_S^2 f^2(\chi)}{\chi^2 \omega_V^2}(S+1) \right]$
  and the final density field is described by Eq. \ref{eq:Gsol}

 \section{MEM Vortex model} \label{app:MEM}
\begin{figure}[h]
\begin{center}
  \includegraphics[width=\linewidth]{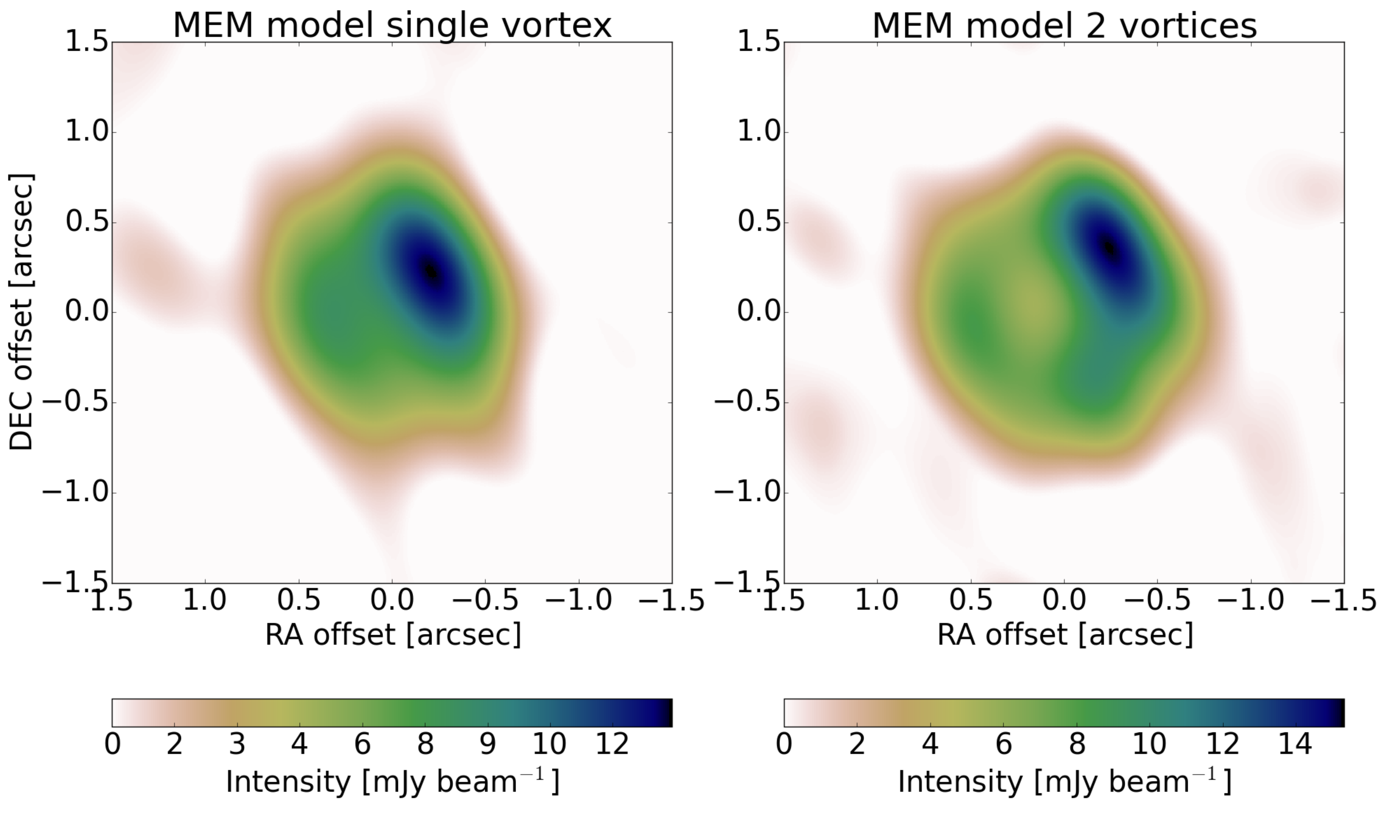}
  \caption{ \label{app:fig_mem} MEM non-parametric models (regularized
    with the maximum entropy method) obtained from simulated
    observations at 337 GHz using the disk model described in
    Sec. \ref{sec:model}. The left panel shows the deconvolved image
    of a disk model with a single vortex, while the right panel shows
    the deconvolved image of a disk model with two vortices. The
    resolution of this observations is approximated by a beam of
    $0.31''\times0.18''$. The $x$ \& $y$ axes indicate the offset from
    the stellar position in RA and DEC in arcsec, i.e. north is up and
    west is right. The stellar position is marked with a red star.}
\end{center}
\end{figure}

 Figure \ref{app:fig_mem} shows the MEM models of the ALMA simulated
 observations with a single and two vortices.

\end{appendix}

\bibliography{SM_pformation.bib}

\end{document}